\documentclass{article}
\usepackage{spconf,amsmath,graphicx}

\usepackage{multirow}
\usepackage{amssymb,amsfonts}
\usepackage{siunitx}
\usepackage{booktabs}  
\usepackage{placeins}
\usepackage{xurl}
\usepackage[breaklinks=true]{hyperref}

\usepackage[T1]{fontenc}
\title{Applicability and Generalization of Audio Fingerprinting Methods to Speech}

\name{Kemal Altwlkany\textsuperscript{1,2}, Elmedin Selmanovic\textsuperscript{2}, Sead Delalić\textsuperscript{2}, Emanuel Lacic\textsuperscript{1}\thanks{This is a preprint version. The peer-reveiwed article is available at:}\thanks{\url{https://doi.org/10.1007/978-3-032-37870-5_28}}}
\address{\textsuperscript{1}Infobip\\\textsuperscript{2}Faculty of Science, University of Sarajevo}

\begin{document}
\ninept
\maketitle
\begin{abstract}
Audio fingerprinting methods identify audio based on its content, even when the content has been possibly corrupted. Despite their applicability and usage in the speech domain, research and advancements in audio fingerprinting traditionally focus on music and are almost exclusively evaluated on music. We provide insights into speech-oriented audio fingerprinting by benchmarking the best performing audio fingerprinting models on speech data while taking into consideration the effects of noise, reverberation, temporal changes, pitch modifications, and lossy compression. Our findings indicate that audio fingerprinting methods based on self-supervised learning outperform classical approaches on speech data, while the performance of these models on speech is inferior compared to the performance that they achieve on music. Results also include an analysis of false positives, showing that fingerprinting methods are more likely to match with an audio of the same speaker rather than an audio of the same text, when mismatches occur.
\end{abstract}


\section{Introduction}

Audio fingerprinting is a widely used technique for identifying audio content, named after the resemblance to how human fingerprints uniquely distinguish every person. Over the years, many different audio fingerprinting methods have been developed sharing a common premise: identify an audio track using only a small segment of the audio. The segment could possibly be degraded by background noise and reverb, lossy compression, or been exposed to time stretching or time contraction.

The pioneering work done at Philips \cite{haitsma2002highly} and Shazam \cite{wang2003industrial} extracted fingerprints from audio by processing it in the spectral domain and extracting distortion-robust features; for example identifying pairs of spectral peaks \cite{wang2003industrial} or analyzing the energy differences of adjacent frequency bands between consecutive spectrogram frames \cite{haitsma2002highly}. This work laid the foundations for further advancements in the field: using wavelets and computer vision techniques \cite{covell2007known,li2015perceptual}, applying different transforms such as the Q-transform, Radon transform, or DCT \cite{li2015perceptual,ozer2005perceptual}, or local feature extraction using SIFT \cite{zhang2015sift}. In addition to research advances, a number of community-driven audio fingerprinting solutions have been made publicly available \cite{audfprint,dejavu,chromaprint}.

The work of Gfeller et al. \cite{gfeller2017now} is widely considered the first audio fingerprinting approach that relied on deep learning, specifically self-supervised contrastive learning \cite{chen2020simple}. Chang et al. \cite{chang2021neural} developed a neural audio fingerprinting method using the same framework as \cite{gfeller2017now,chen2020simple}, but with a convolutional encoder. In recent years, given the rapid advances in AI, approaches relying on more advanced neural architectures such as the transformer \cite{singh2022attention,singh2023simultaneously}, the conformer \cite{altwlkany2025pretrained,gulati20interspeech} or graph neural networks \cite{bhattacharjee2025grafprint} have appeared. Most of these approaches generate dense low-dimensional embeddings and instead of hashing rely on vector search, most notably, by using the vector database Faiss \cite{johnson2019billion}. These approaches (models) have pushed the state-of-the-art far beyond earlier benchmarks. For example, in \cite{haitsma2002highly} and \cite{ozer2005perceptual}, audio is degraded with white gaussian noise from \SI{20}{\dB} to \SI{35}{\dB} SNR and time speedup/slowdown of 2\% to 6\% is applied, while in newer approaches these values are benchmarked at \SI{0}{\dB} SNR non-static background noise or extreme $\pm 50$\% time speedup/slowdown \cite{altwlkany2025pretrained,bhattacharjee2025grafprint,cortes2025peaknetfp}.

Audio fingerprinting has been successfully applied in commercial products such as audio recognition systems \cite{haitsma2002highly,wang2003industrial,covell2007known,gfeller2017now,chang2021neural}, for detecting copyright infringements \cite{wang2003industrial,chang2021neural,altwlkany2025pretrained}, but also in telecommunications to de-duplicate audio and quickly identify announcements played during session initiation \cite{altwlkany2025pretrained,prasad2020s}.

\subsection{Speech fingerprinting}

Despite being called \textit{audio} fingerprinting methods, most methods are almost exclusively evaluated using \textit{music} datasets. Classical work on audio fingerprinting was motivated by music identification services \cite{haitsma2002highly,wang2003industrial,covell2007known,zhang2015sift,gfeller2017now}, while most modern self-supervised approaches \cite{chang2021neural,singh2022attention,singh2023simultaneously,altwlkany2025pretrained,bhattacharjee2025grafprint,cortes2025peaknetfp} were trained on the FMA dataset \cite{defferrard2016fma}. Performance of novel audio fingerprinting methods is not evaluated on audio content other than music.

Although scarce, there is some work on the topic of specifically \textit{speech} fingerprinting. In \cite{altwlkany2024application}, the authors present modifications of Waveprint \cite{covell2007known} which they apply for speech retrieval. Renza et al. \cite{renza2019robust} developed an algorithm for digital forensics; they extract MFCCs from audio and further reduce dimensionality by applying PCA. In \cite{ozer2005perceptual}, the authors evaluate three perceptual audio hash functions on speech and music. Two of the hash functions perform slightly better on music compared to speech.

We argue that speech fingerprinting has its importance, e.g., in digital forensics, but also in the aforementioned use cases in telephony or speech retrieval \cite{altwlkany2025pretrained,prasad2020s,renza2019robust}. Modern audio fingerprinting methods based on self-supervised learning have set the state-of-the-art for audio retrieval, but are benchmarked on music only. Through a series of 25 different experiments, we first analyze the performance of audio fingerprinting methods on speech data based on the type of possible audio distortion: background noise, reverb, temporal distortions, pitch modifications, or lossy compression. Second, we assess the disparity between the performances of these methods when evaluated on speech and music. Finally, we provide deeper insights into the behavior of speech fingerprinting by analyzing false positives, i.e. if a fingerprinting method matches with fingerprints that do not belong to the same file, is there a specific pattern that can be observed from these mismatches? As we demonstrate in the results, if the correct audio has not been found, audio fingerprinting methods tend to match with other files that share the same speaker, rather than the same text. Our findings indicate that although state-of-the-art audio fingerprinting methods exhibit good speech retrieval rates, the results are far from their music counterparts, especially in conditions with severe audio distortions.

\section{Method}

This section provides an overview of the selected models, the data, and the overall experimental setup.

\subsection{Models}

As a baseline, we select a reliable and community-driven classical audio fingerprinting method - \textit{audfprint}\cite{audfprint}. For self-supervised representatives, we select two modern approaches, each featuring a distinct architecture; pretrained conformers (PTC) \cite{altwlkany2025pretrained} - conformer encoder, and GraFPrint \cite{bhattacharjee2025grafprint} - GNN encoder. These approaches are novel and achieve state-of-the-art results for audio retrieval tasks across different types of audio distortions. Additionally, the authors of both methods made checkpoint models publicly available, which is a requirement of ours, as we wish to avoid retraining the models ourselves so that our setup remains as close to the original setup as possible.

\subsection{Vector search}
Self-supervised audio fingerprinting techniques generate fingerprints for audio in the form of embeddings, which are stored and retrieved using vector databases.
Faiss \cite{johnson2019billion} is the most commonly used vector database by self-supervised methods \cite{chang2021neural,singh2022attention,singh2023simultaneously,altwlkany2025pretrained,bhattacharjee2025grafprint,cortes2025peaknetfp}. It supports many advanced indexing mechanisms, providing a trade-off between speed and accuracy. As our goal is to evaluate the overall accuracy of fingerprinting methods for speech, we use a naive, exhaustive search in Faiss, thereby guaranteeing that if a match exists, it will be found. Similarly, all mismatches are attributed to the underlying models and not to Faiss, thereby eliminating the need to consider the effects of using Faiss in our analysis.

\subsection{Data}
We use the VCTK speech corpus \cite{yamagishi2019vctk}, which has slightly less than 45 thousand audio files. A drawback of this dataset is that it is in English language only. The benefits are that it features high-quality audio sampled at \SI{48}{\kilo\Hz} and a set of 110 speakers of diverse accents and gender. Another advantage is the availability of transcripts, which will enable us to compare textual retrieval rates as well (e.g. whether retrieved files match by text if not otherwise).

Audio distortion is achieved by applying audio augmentations through the Python audiomentations package \cite{iver2025zenodo}.  We use the test split of the MS-SNSD dataset \cite{reddy19_interspeech} for background noise. Since the fingerprinting methods were trained on different room impulse response datasets, we could not easily find a dataset that would not favour a certain method, so we decided to synthesize room impulse responses artificially through \cite{iver2025zenodo}. The rest of the augmentations such as temporal distortions, pitch shifts, or compression do not require data.

\subsection{Procedure}
Regarding Audfprint, it manages its own storage, so we simply fingerprint the VCTK corpus using audfprint and add the fingerprints into the storage. For GraFPrint and PTC, we create a separate Faiss vector database for each of them and populate the databases with the computed fingerprints of the VCTK corpus. Measuring the performance of a method is done by taking the query (augmented) sample, computing its fingerprints, and searching for the best match within the database.

\subsection{Experiments}

In total, 25 experiments were performed. The experiments were varied by type and intensity of audio distortion: adding noise at different SNR levels, stretching or contracting the audio in time, pitch shifting the audio, applying reverb, etc. In each experiment, we randomly select 10 files for each of the 110 speakers, totaling 1,100 files per experiment. All of these files were augmented and saved to disk before being processed further by any of the fingerprinting methods. This was to ensure that, during each experiment, all fingerprinting methods were shown the same augmented versions of the files.

The target benchmark (metric) in audio fingerprinting is hit rate: given a set of query files, what percentage of fingerprints from those files were correctly retrieved, that is, how many fingerprints were correctly matched with their corresponding fingerprint from the database \cite{chang2021neural,singh2022attention,singh2023simultaneously,altwlkany2025pretrained,bhattacharjee2025grafprint,cortes2025peaknetfp}. In this context, the metric is also referred to as accuracy or retrieval rate. All benchmarks that we report were computed at a 95\% confidence interval using the bootstrap approach \cite{tibshirani1993introduction}.

\section{Results and discussion}
We briefly consider the time and memory complexity of fingerprinting methods.

\begin{figure}[t]
  \centering
  \includegraphics[width=0.75\linewidth]{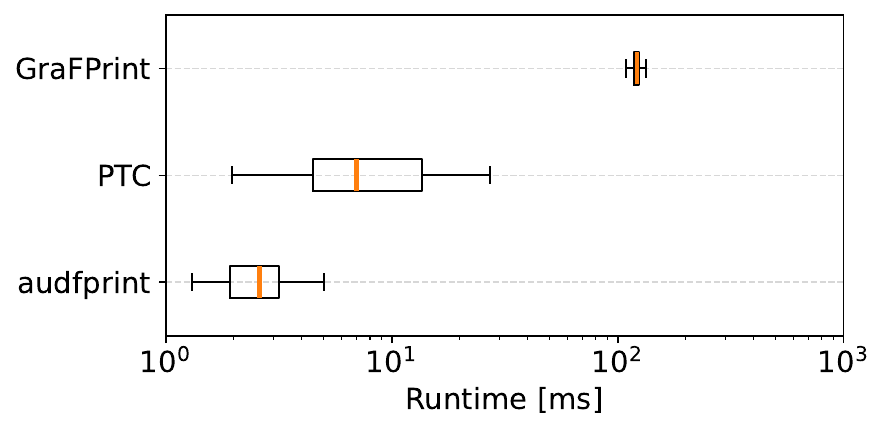}
  \caption{Runtime of fingerprinting methods shown at logarithmic scale. The orange line indicates the median value.}
  \label{fig:runtime-boxplots}
\end{figure}

The distribution of runtimes is shown in Figure \ref{fig:runtime-boxplots}. These values are normalized per unit of input audio, i.e. they show how much time is required to process 1 second of input audio. audfprint is faster by an order of magnitude (mean: $2.74 \pm 0.11$, median: $2.60$ \SI{}{\milli\second}) compared to PTC (mean: $13.72 \pm 2.23$, median: $6.98$ \SI{}{\milli\second}) and GraFPrint (mean: $129.53 \pm 2.89$, median: $121.821$ \SI{}{\milli\second}), but PTC executes faster than GraFPrint by an order of magnitude. The results were benchmarked using an Apple Mac M4 Max.

Table \ref{tab:memory-footprint} shows the number of fingerprints generated for the entire VCTK speech corpus per fingerprinting method, as well as the number of bits required for storing a single fingerprint. Most fingerprints were generated by audfprint, but it consumes less space since its fingerprints are lightweight hashes and not dense floating-point embeddings. GraFPrint requires 2.92 times more storage than PTC. This result is expected, as the stride (difference) between two consecutive fingerprints in PTC is \SI{300}{\milli\second}, compared to \SI{100}{\milli\second} in GraFPrint.

\begin{table}[t]
    \centering
    \caption{Memory footprint per fingerprinting method for storing the fingerprints of VCTK speech corpus without metadata. Note that for audfprint a single hash is considered a fingerprint.}
    \resizebox{1.0\linewidth}{!}{%
\centering

\begin{tabular}{l|rrr}
          & \multicolumn{1}{l}{fingerprints} & \multicolumn{1}{l}{fingerprint size (bits)} & \multicolumn{1}{l}{total size (MB)}  \\ 
\hline
audfprint & 1,509,822                        & 20                                          & 3.59                             \\
GraFPrint & 1,107,864                        & 4096                                        & 540                                  \\
PTC       & 378,248                          & 4096                                        &     184                             
\end{tabular}

    }
    \label{tab:memory-footprint}
\end{table}

\begin{table}[t]
    \centering
    \caption{Retrieval rates when the query sample has been distorted by artificially synthesized reverbant conditions.}
    \resizebox{1.0\linewidth}{!}{%
\centering

\begin{tabular}{l|lll}
              & audfprint                 & GraFPrint                 & PTC                        \\ 
\hline
Hit rate (\%) & \multicolumn{1}{r}{$27.82 \pm 5.18$} & \multicolumn{1}{r}{$77.21 \pm 4.88$} & \multicolumn{1}{r}{$47.56 \pm 5.56$} 
\end{tabular}

    }
    \label{tab:reverb}
\end{table}

\begin{figure*}[t]
  \centering
  \includegraphics[width=\linewidth]{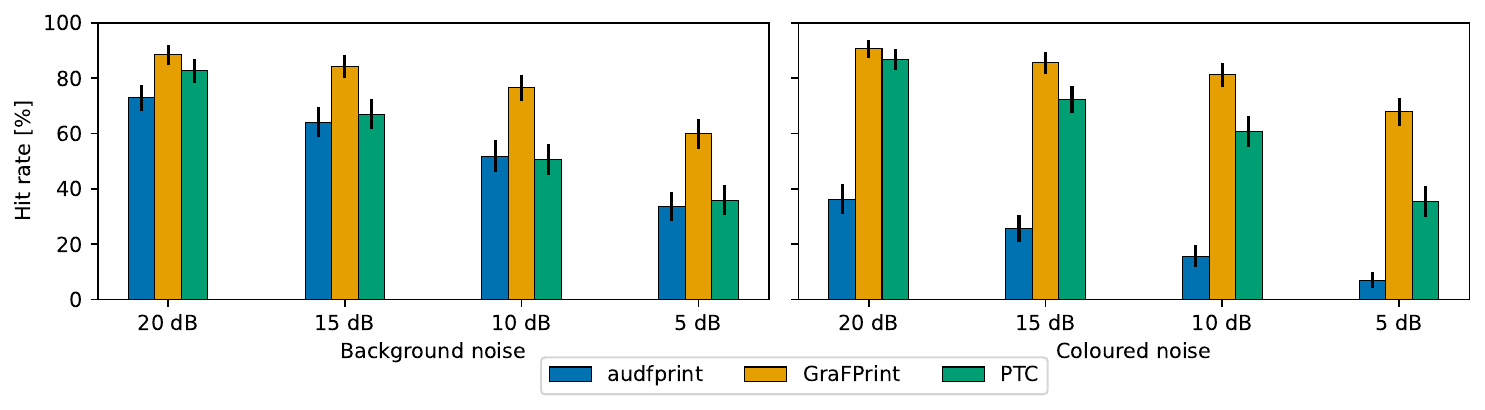}
  \caption{Retrieval rates when the query sample has been corrupted by background noise or coloured noise.}
  \label{fig:noise-experiments}
\end{figure*}

\subsection{Noise and reverb}

Our first experiments involved adding background noise and coloured noise at various SNR levels to the query samples. Coloured noise was synthesized programmatically, while we used the test split of the MS-SNSD dataset \cite{reddy19_interspeech} for background noise. The results are represented graphically in Figure \ref{fig:noise-experiments}. GraFPrint achieves the highest hit rates, especially during low SNR conditions. Both GraFPrint and PTC consistently outperform audfprint, especially for coloured noise. The hit rates when the query sample is exposed to reverb are shown in Table \ref{tab:reverb}. Overall, the accuracies which are achieved here are lower compared to the accuracies which these fingerprinting methods achieve when benchmarked on music datasets \cite{altwlkany2025pretrained,bhattacharjee2025grafprint}, but the performance of GraFPrint is significantly better than that of audfprint and PTC.

\begin{table}[t]
    \centering
    \caption{Retrieval rates when the query sample has been temporally offset from the reference sample.}
    \resizebox{1.0\linewidth}{!}{%
\centering

\begin{tabular}{l|lll}
              & audfprint                 & GraFPrint                 & PTC                        \\ 
\hline
Hit rate (\%) & \multicolumn{1}{r}{$91.00 \pm 3.33$} & \multicolumn{1}{r}{$80.89 \pm 4.55$} & \multicolumn{1}{r}{$96.69 \pm 2.02$}
\end{tabular}

    }
    \label{tab:temporal-shift}
\end{table}

\begin{figure}[t]
  \centering
  \includegraphics[width=\linewidth]{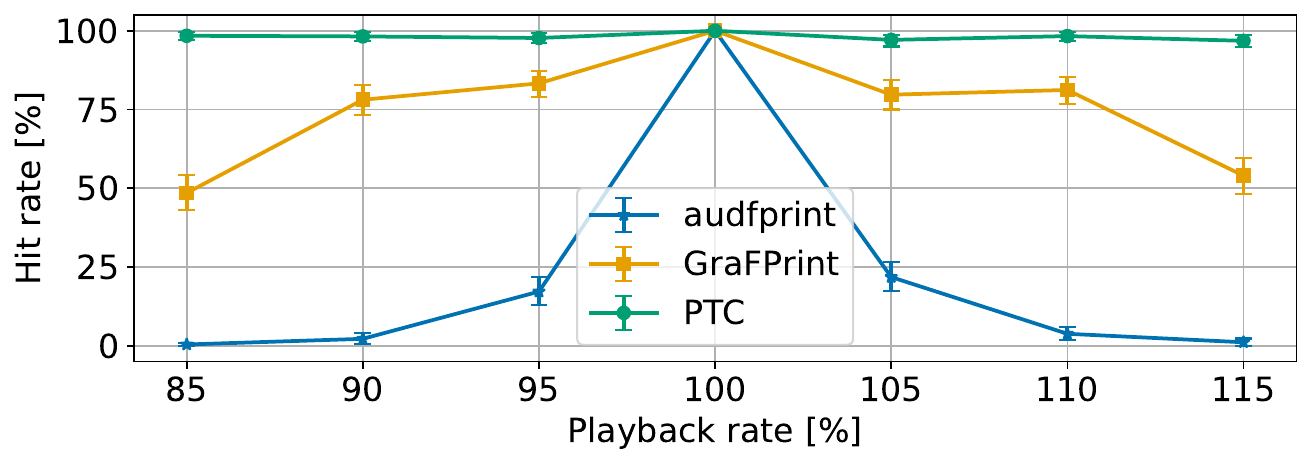}
  \caption{Retrieval rates when the query sample's playback rate has been modified, resulting in either a sped up or slowed down version of the audio.}
  \label{fig:temporal-experiments}
\end{figure}

\begin{table}
    \centering
    \caption{Retrieval rates when the query sample has been pitch-shifted.}
    \resizebox{1.0\linewidth}{!}{%
\centering

\begin{tabular}{r|rrr}
                        & \multicolumn{1}{c}{audfprint} & \multicolumn{1}{c}{GraFPrint} & \multicolumn{1}{c}{PTC}  \\ 
\hline
Downshift 1 semitone    & $0.00 \pm 0.00$                          & $97.26 \pm 1.94$                         & $52.84 \pm 5.82$                    \\
Downshift 0.5 semitones & $0.09 \pm 0.58$                          & $97.87 \pm 1.87$                         & $93.11 \pm 3.11$                    \\
Upshift 0.5 semitones   & $0.08 \pm 0.55$                          & $98.10 \pm 1.94$                         & $95.04 \pm 2.71$                    \\
Upshift 1 semitone      & $0.00 \pm 0.00$                          & $97.98 \pm 1.65$                         & $41.98 \pm 5.69$                   
\end{tabular}

    }
    \label{tab:pitch-shift}
\end{table}

\subsection{Temporal misalignments}

We analyzed two types of temporal misalignments. For the first, we applied a random temporal shift (also referred to as time/temporal offset or stride), which means starting the fingerprinting process from a randomly selected point in time within the given audio file. The second form of temporal misalignment is applying either temporal stretching or contraction, which means speeding up or slowing down the audio. The results are summarized in Table \ref{tab:temporal-shift} and visualized in Figure \ref{fig:temporal-experiments}.

In terms of temporal shift, GraFPrint provides the lowest accuracy. This is a reasonable outcome as GraFPrint creates fingerprints from very small segments of audio, so slight alterations can have strong effects. Both deep learning approaches outperform audfprint in cases of temporal stretching and contraction, with PTC maintaining near 100\% hit rate even in cases of 15\% stretch or contraction. These results are in line with the results obtained for music datasets, since PTC claim to attain near 100\% accuracy for extreme cases of temporal stretch/contraction \cite{altwlkany2025pretrained}.

\subsection{Pitch modifications}

Pitch shifting is a commonly used effect in speech processing that modifies the fundamental frequency (F0) by scaling it up or down while typically preserving the temporal structure of the signal. Results are summarized in Table \ref{tab:pitch-shift}. GraFPrint has a slight advantage over PTC at half a semitone, but at one semitone the performance of PTC drops significantly. This could be caused by PTC not being trained on enough samples with pitch distortions. Pitch shifting proved to be an incredibly difficult task for audfprint, which is not surprising given the nature of the fingerprinting method (locating features within the spectrogram, which became completely altered due to changes in pitch). We also note that the uncertainty around the confidence interval for PTC is greater than that of GraFPrint, for example, at 1 semitone downshift the CI for PTC spans more than 10\% of the mean value.

\begin{figure}[t]
  \centering
  \includegraphics[width=0.9\linewidth]{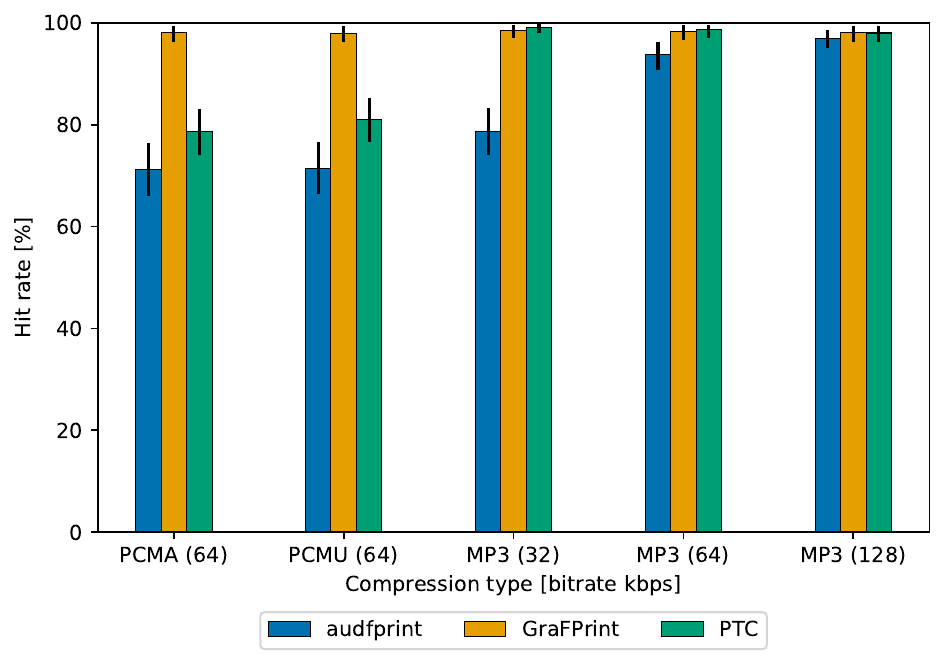}
  \caption{Retrieval rates of fingerprinting methods when the query sample has been quantized and downsampled by a lossy compression format or codec.}
  \label{fig:compression-experiments}
\end{figure}

\subsection{Effects of lossy audio compression}

One of the most popular lossy compression algorithms is MP3 (formally MPEG-1 Audio Layer III). Additionally, since we found applications of audio fingerprinting in telephony \cite{altwlkany2025pretrained,prasad2020s}, we compress/transcode audio using the official standard telephony codec defined by the ITU-T which is G.711 (both $A$-law (PCMA) and $\mu$-law (PCMU) variants) \cite{g711}.

Results are shown in Figure \ref{fig:compression-experiments}. Unlike previous experiments, the differences are not clear immediately, especially for MP3, so we performed Pearson's chi-square test for independence. At 32kbps MP3 compression results indicated a significant difference between the models, $\chi^2(2,\,\, N=40\,074) = 2231, p < .001$. The association between model and retrieval rate was moderate, as indicated by Cramer’s $V = 0.236$. Post-hoc pairwise comparisons were performed using chi-square tests with a Bonferroni correction applied for multiple comparisons (adjusted $\alpha = .0167$). All pairwise differences were statistically significant ($p < .001$).

At 64kbps MP3 compression, we obtained a significant result, $\chi^2(2,\,\, N=38\,739) = 145, p < .001$, but the association is weak: Cramer’s $V = 0.061$. Post-hoc pairwise comparisons with Bonferroni correction showed no statistically significant difference between GraFPrint and PTC. Finally, at 128kbps, we obtain $\chi^2(2,\,\, N=38\,695) = 6.01, p < .001$, with Cramer’s $V = 0.0125$. Despite being statistically significant, the difference is negligible in terms of effect size. Furthermore, post-hoc comparisons did not show statistically significant differences between any of the pairs ($p > .0167$).
Results are shown in Figure \ref{fig:compression-experiments}. All fingerprinting methods perform well on MP3 compressed audio, with PTC having a very slight advantage over GraFPrint. The performance of audfprint is consistently below that of GraFPrint and PTC, but the difference is smaller than that of pitch-shifted audio. For PCMA/PCMU compressed audio, GraFPrint has the best performance. This shows that the method is capable of generalizing well to non-linear quantization effects, since it was not trained on this type of degradation.

\subsection{Analysis of mismatches}

We were interested in whether there is a specific pattern in the mismatches that happen, i.e., in case that a fingerprinting method fails to retrieve the exact audio file we want to investigate whether there are specific patterns in the audio files that it actually retrieved (matched with). The analysis is limited to GraFPrint and PTC, since audfprint operates in a different fashion and does not provide access to mismatch data, i.e. we don't know which stored embeddings were matched to which query embeddings. We specifically analyze the instances where the two methods have the largest drop in performance, which are the temporal contraction at 15\%, the single semitone pitch shift, and reverb. Roughly 50\% of the queries were mismatches in these three experiments by either PTC or GraFPrint.

\begin{figure}[t]
  \centering
  \includegraphics[width=\linewidth]{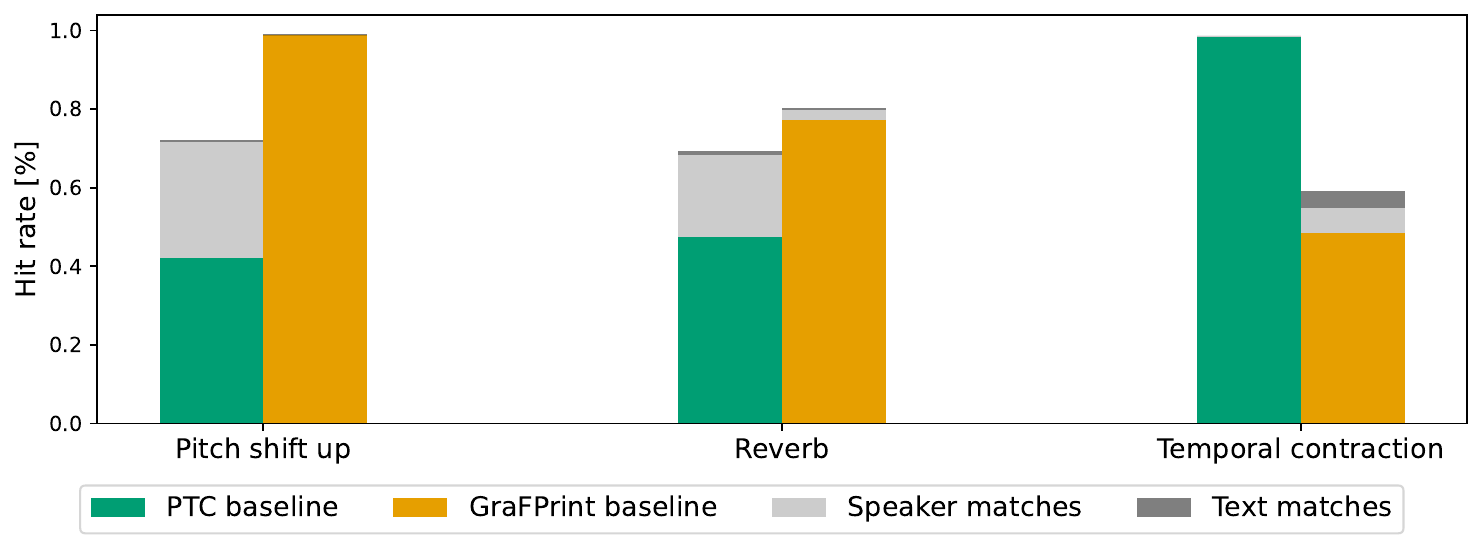}
  \caption{Mismatch analysis: fingerprinting methods are more likely to match with the same speaker, rather than the same text.}
  \label{fig:speaker-and-text-performance}
\end{figure}

If we break down the mismatches by speaker and textual information, we observe a relationship between falsely matched embeddings and query embeddings. These are visualized in Figure \ref{fig:speaker-and-text-performance}, which shows the retrieval rates, but additionally includes how many of the mismatches actually belong to the correct speaker, or how many of them contain the same text.

29.7\% of mismatches produced by PTC for pitch shifted audio belong to the correct speaker, which means that the method correctly identified the speaker, but not the spoken content. Similarly, 20.8\% of mistmaches belong to the same speaker during reverbant conditions. On the other hand, for pitch shifted audio none of the mismatches match by text, i.e. the wrongly retrieved embedding does not share the same text as the query embedding. This is true during reverb as well, with text being correctly matched in just 0.08\% of mismatches.

GraFPrint's  worst results were benchmarked at 15\% temporal contraction. Out of all mismatches, 6.4\% belong to the correct speaker, while 4.2\% contain the same text (spoken content). Unlike PTC, in GraFPrint the mismatches are less correlated with the query embedding, with only a total of 10.6\% of mismatched embeddings being correlated to the query embedding by either speaker or text, while these rates were between 20 and 30\% for PTC.

\begin{table}[t]
    \centering
    \caption{Comparison of audio fingerprinting retrieval rates between music and speech data.}
    \resizebox{1.0\linewidth}{!}{%
\centering

\begin{tabular}{cl|r|r|r|r}
\multicolumn{1}{l}{}    & \multicolumn{1}{l}{} & \multicolumn{4}{c}{Background noise}                                                                            \\
\multicolumn{1}{l}{}    &                      & \multicolumn{1}{l|}{5 dB} & \multicolumn{1}{l|}{10 dB} & \multicolumn{1}{l|}{15 dB} & \multicolumn{1}{l}{20dB}  \\ 
\hline
\multirow{3}{*}{Speech} & audfprint          & $33.6 \pm 5.4$                     & $51.7 \pm 5.7$                      & $64.2 \pm 5.5$                     & $72.9 \pm 5.1$                     \\
                        & GraFPrint            & $59.9 \pm 5.6$                     & $76.5 \pm 4.8$                      & $84.4 \pm 4.1$                       & $88.4 \pm 3.8$                     \\
                        & PTC           & $35.9 \pm 5.4$                     & $50.7 \pm 5.7$                     & $67.1 \pm 5.4$                      & $82.9 \pm 4.2$                     \\ 
\hline
\multirow{3}{*}{Music}  & audfprint           & $63.4 \pm 5.4$                     & $76.2 \pm 4.9$                      & $85.3 \pm 4.3$                      & $89.1 \pm 3.8$                      \\
                        & GraFPrint \cite{bhattacharjee2025grafprint}            & 98.3                      & 99.3                       & 99.6                       & 99.9                      \\
                        & PTC \cite{altwlkany2025pretrained}           & 93.9                      & 97.1                       & 97.8                       & /                        
\end{tabular}
    }
    \label{tab:music-vs-speech}
\end{table}


\subsection{Comparison with performance on music}

Table \ref{tab:music-vs-speech} provides the retrieval rates of audfprint, GraFPrint and PTC when benchmarked on speech and music data. For speech data, the results were computed on the VCTK dataset \cite{yamagishi2019vctk}. For music data, we referred to the results reported for GraFPrint \cite{bhattacharjee2025grafprint} and PTC \cite{altwlkany2025pretrained} performed on the FMA dataset \cite{defferrard2016fma}.

The results show that all methods perform better on music data regardless of SNR value. However, the gap between music and speech retrieval rates is more prominent as the SNR values decrease; for example, at \SI{5}{\dB} there is an absolute difference of 60.35\% in retrieval rate for PTC \cite{altwlkany2025pretrained}. GraFPrint at \SI{20}{\dB} achieves near-perfect results on music - 99.9\% \cite{bhattacharjee2025grafprint}, but has an accuracy of 88.4\% on speech.

\section{Applicability and Discussion}

It is worth discussing the applicability of speech fingerprinting in practice. Automatic speech recognition is the natural choice to determine the spoken content of a recording. However, if the collection of recordings is known in advance, then speech fingerprinting would offer several benefits; it is computationally more efficient than automatic speech recognition, it's performance is independent of the language of the recording, and it requires only a few seconds of the original recording to perform identification. However, this approach does have a drawback if the collection of audio files contains multiple recordings of the same speaker with overlapping content (e.g. utterances of the same phrase). In that case, the minimum duration of the query sample must be adequately increased to eliminate ambiguous matches.

One might argue about the desired outcome of speech fingerprinting in cases of specific audio distortions, such as pitch modifications; should an audio fingerprinting model match with same speaker even though the pitch was strongly modified? Perhaps it would make more sense to match with another speaker that sounds more similar to the original audio and has similar spoken content.

Given our analysis, we suggest that the results obtained during benchmarking of models on music data are not observed on speech data. In general, all methods perform worse on speech data compared to music data. GraFPrint is a well-rounded method that offers the best performance in all scenarios except for MP3 compression and temporal distortions, so it is the best choice. Should temporal distortions be expected, PTC would be the obvious choice. The advantage of audfprint is the low memory requirement and faster runtime, which could be beneficial on devices with low computational power.

\section{Conclusion}
In this paper we provided a thorough analysis of audio fingerprinting methods when applied to speech. We showed which methods performed well on different types of audio distortions that are common when applying audio fingerprinting in practice. Methods based on self-supervised learning have shown to generalize better and perform better than classical audio fingerprinting approaches. Despite that, their performance on speech is still sub-par compared to the results that they achieve on music. 
Results also showed that when dealing with speech, audio fingerprinting methods are more likely to match the same speaker, rather than the same text. We plan to continue our work by fine-tuning or adapting these approaches so that their performances for speech match their music counterpart.

\section{Acknowledgments}
This research was supported in part by the project Infobip Global Communication Platform (PK.1.1.07.0001), part of the Important Project of Common European Interest on Next Generation Cloud Infrastructure and Services (IPCEI-CIS) consortium.

\bibliographystyle{IEEEbib}
\bibliography{references}

@inproceedings{wang2003industrial,
  title={An industrial strength audio search algorithm.},
  author={Wang, Avery and others},
  booktitle={Proc. of the Int. Society for Music Information Retrieval (ISMIR)},
  volume={4},
  pages={7--13},
  year={2003},
}

@inproceedings{haitsma2002highly,
 title={A highly robust audio fingerprinting system.},
 author={Haitsma, Jaap and Kalker, Ton},
 volume={3},
 year={2002},
 pages={107--115},
 booktitle={Proc. of the Int. Society for Music Information Retrieval (ISMIR)}
}

@INPROCEEDINGS{covell2007known,
  author={Covell, Michele and Baluja, Shumeet},
  booktitle={2007 IEEE International Conference on Acoustics, Speech and Signal Processing - ICASSP '07}, 
  title={Known-Audio Detection using Waveprint: Spectrogram Fingerprinting by Wavelet Hashing}, 
  year={2007},
  volume={1},
  number={},
  pages={I-237-I-240},
  doi={10.1109/ICASSP.2007.366660}}

@article{zhang2015sift,
  title={{SIFT-based local spectrogram image descriptor: a novel feature for robust music identification}},
  author={Zhang, Xiu and Zhu, Bilei and Li, Linwei and Li, Wei and Li, Xiaoqiang and Wang, Wei and Lu, Peizhong and Zhang, Wenqiang},
  journal={EURASIP Journal on Audio, Speech, and Music Processing},
  volume={2015},
  number={1},
  pages={6},
  year={2015},
  publisher={Springer}
}

@article{ozer2005perceptual,
  title={Perceptual audio hashing functions},
  author={{\"O}zer, Hamza and Sankur, B{\"u}lent and Memon, Nasir and Anar{\i}m, Emin},
  journal={EURASIP Journal on Advances in Signal Processing},
  volume={2005},
  number={12},
  pages={658950},
  year={2005},
  publisher={Springer}
}

@inproceedings{li2015perceptual,
  title={{Perceptual audio hashing using RT and DCT in wavelet domain}},
  author={Li, Jinfeng and Wu, Tao},
  booktitle={2015 11th International Conference on Computational Intelligence and Security (CIS)},
  pages={363--366},
  year={2015},
  organization={IEEE}
}

@inproceedings{gfeller2017now,
  title={Now playing: Continuous low-power music recognition},
  author={Gfeller, Beat and Guo, Ruiqi and Kilgour, Kevin and Kumar, Sanjiv and Lyon, James and Odell, Julian and Ritter, Marvin and Roblek, Dominik and Sharifi, Matthew and Velimirovi{\'c}, Mihajlo and others},
  booktitle={NeurIPS 2017 Workshop on Machine Learning on the Phone and other Consumer Devices},
  year={2017}
}

@inproceedings{chen2020simple,
  title={A simple framework for contrastive learning of visual representations},
  author={Chen, Ting and Kornblith, Simon and Norouzi, Mohammad and Hinton, Geoffrey},
  booktitle={International conference on machine learning},
  pages={1597--1607},
  year={2020},
  organization={PmLR}
}

@inproceedings{chang2021neural,
  title={Neural audio fingerprint for high-specific audio retrieval based on contrastive learning},
  author={Chang, Sungkyun and Lee, Donmoon and Park, Jeongsoo and Lim, Hyungui and Lee, Kyogu and Ko, Karam and Han, Yoonchang},
  booktitle={IEEE International Conference on Acoustics, Speech and Signal Processing (ICASSP 2021)},
  pages={3025--3029},
  year={2021},
  organization={IEEE}
}

@inproceedings{singh2022attention,
  title={Attention-based audio embeddings for query-by-example},
  author={Singh, Anup and Demuynck, Kris and Arora, Vipul},
  booktitle={Proc. of the Int. Society for Music Information Retrieval (ISMIR)},
  volume={23},
  pages={52--58},
  year={2022},
}

@INPROCEEDINGS{singh2023simultaneously,
  author={Singh, Anup and Demuynck, Kris and Arora, Vipul},
  booktitle={ICASSP 2023 - 2023 IEEE International Conference on Acoustics, Speech and Signal Processing (ICASSP)}, 
  title={Simultaneously Learning Robust Audio Embeddings and Balanced Hash Codes for Query-by-Example}, 
  year={2023},
  volume={},
  number={},
  pages={1-5},
  doi={10.1109/ICASSP49357.2023.10096103}}

@inproceedings{cortes2025peaknetfp,
  title        = {PeakNetFP: Peak-Based Neural Audio Fingerprinting Robust to Extreme
                  Time Stretching},
  author       = {Guillem Cort{\`{e}}s{-}Sebasti{\`{a}} and
                  Benjamin Martin and
                  Emilio Molina and
                  Xavier Serra and
                  Romain Hennequin},
 volume={26},
 year={2025},
 pages={206--214},
 booktitle={Proc. of the Int. Society for Music Information Retrieval (ISMIR)}
}

@article{altwlkany2025pretrained,
  title={Pretrained Conformers for Audio Fingerprinting and Retrieval},
  author={Altwlkany, Kemal and Selmanovic, Elmedin and Delalic, Sead},
  journal={arXiv preprint arXiv:2508.11609},
  year={2025}
}

@inproceedings{bhattacharjee2025grafprint,
  title={{Grafprint: A GNN-based approach for audio identification}},
  author={Bhattacharjee, Aditya and Singh, Shubhr and Benetos, Emmanouil},
  booktitle={ICASSP 2025-2025 IEEE International Conference on Acoustics, Speech and Signal Processing (ICASSP)},
  pages={1--5},
  year={2025},
  organization={IEEE}
}

@article{johnson2019billion,
  title={Billion-scale similarity search with {GPUs}},
  author={Johnson, Jeff and Douze, Matthijs and J{\'e}gou, Herv{\'e}},
  journal={IEEE Transactions on Big Data},
  volume={7},
  number={3},
  pages={535--547},
  year={2019},
  publisher={IEEE}
}

@inproceedings{prasad2020s,
  title={Who's calling? characterizing robocalls through audio and metadata analysis},
  author={Prasad, Sathvik and Bouma-Sims, Elijah and Mylappan, Athishay Kiran and Reaves, Bradley},
  booktitle={29th USENIX Security Symposium (USENIX Security 20)},
  pages={397--414},
  year={2020}
}

@inproceedings{defferrard2016fma,
  title={FMA: A dataset for music analysis},
  author={Defferrard, Micha{\"e}l and Benzi, Kirell and Vandergheynst, Pierre and Bresson, Xavier},
  booktitle={Proc. of the Int. Society for Music Information Retrieval (ISMIR)},
  volume={18},
  pages={316--323},
  year={2017},
}

@inproceedings{altwlkany2024application,
  title={Application of Audio Fingerprinting Techniques for Real-Time Scalable Speech Retrieval and Speech Clusterization},
  author={Altwlkany, Kemal and Delali{\'c}, Sead and Alihod{\v{z}}i{\'c}, Adis and Selmanovi{\'c}, Elmedin and Hasi{\'c}, Damir},
  booktitle={2024 47th MIPRO ICT and Electronics Convention (MIPRO)},
  pages={91--96},
  year={2024},
  organization={IEEE}
}

@article{renza2019robust,
  title={Robust speech hashing for digital audio forensics},
  author={Renza, Diego and Vargas, Jaisson and Ballesteros, Dora M},
  journal={Applied Sciences},
  volume={10},
  number={1},
  pages={249},
  year={2019},
  publisher={MDPI}
}

@misc{yamagishi2019vctk,
  author       = {Yamagishi, Junichi and Veaux, Christophe and MacDonald, Kirsten},
  title        = {{CSTR VCTK Corpus: English Multi-speaker Corpus for CSTR Voice Cloning Toolkit}},
  year         = {2019},
  version      = {0.92},
  note = {doi: doi.org/10.7488/ds/2645},
  institution  = {University of Edinburgh, The Centre for Speech Technology Research (CSTR)}
}

@inproceedings{reddy19_interspeech,
  title     = {A Scalable Noisy Speech Dataset and Online Subjective Test Framework},
  author    = {Chandan K.A. Reddy and Ebrahim Beyrami and Jamie Pool and Ross Cutler and Sriram Srinivasan and Johannes Gehrke},
  year      = {2019},
  booktitle = {Interspeech 2019},
  pages     = {1816--1820},
  doi       = {10.21437/Interspeech.2019-3087},
  issn      = {2958-1796},
}

@misc{iver2025zenodo,
    author =  {{Iver Jordal}},
    title = {{“iver56/audiomentations: v0.42.0”}},
    howpublished = { Zenodo},
    year = {2025},
    note = {doi: 10.5281/zenodo.15806235}
}

@article{tibshirani1993introduction,
  title={An introduction to the bootstrap},
  author={Tibshirani, Robert J and Efron, Bradley},
  journal={Monographs on statistics and applied probability},
  volume={57},
  number={1},
  pages={1--436},
  year={1993}
}

@misc{audfprint,
  author       = {Dan Ellis},
  title        = {audfprint},
  year         = {2014},
  howpublished = {\url{https://github.com/dpwe/audfprint}},
  note         = {{Landmark-based audio fingerprinting. GitHub repository [online].}}
}

@misc{dejavu,
  author       = {Will Drevo},
  title        = {dejavu},
  year         = {2013},
  howpublished = {\url{https://github.com/worldveil/dejavu}},
  note         = {{Audio fingerprinting and recognition in Python. GitHub repository [online].}}
}

@misc{chromaprint,
  author       = {AcousticID},
  title        = {{Chromaprint}},
  year         = {2010},
  howpublished = {\url{https://github.com/acoustid/chromaprint}},
  note         = {{C library for generating audio fingerprints. GitHub repository [online].}}
}

@inproceedings{gulati20interspeech,
  title     = {Conformer: Convolution-augmented Transformer for Speech Recognition},
  author    = {Anmol Gulati and James Qin and Chung-Cheng Chiu and Niki Parmar and Yu Zhang and Jiahui Yu and Wei Han and Shibo Wang and Zhengdong Zhang and Yonghui Wu and Ruoming Pang},
  year      = {2020},
  booktitle = {Interspeech 2020},
  pages     = {5036--5040},
  doi       = {10.21437/Interspeech.2020-3015},
  issn      = {2958-1796},
}

@techreport{g711,
  author      = {{International Telecommunication Union}},
  title       = {Pulse code modulation ({PCM}) of voice frequencies},
  type        = {ITU-T Recommendation},
  number      = {G.711},
  institution = {International Telecommunication Union},
  address     = {Geneva, Switzerland},
  month       = nov,
  year        = {1988},
  url         = {https://www.itu.int/rec/T-REC-G.711-198811-I/en}
}

\end{document}